# Sub-cycle Electron Dynamics in the Generation of Below Threshold Harmonics

Li Wang[*123], Jinxing Xue[*123], Zhinan Zeng[#123], Ruxin Li[†123] and Zhizhan Xu[123]

[1]*State Key Laboratory of High Field Laser Physics, Shanghai Institute of Optics and Fine Mechanics, Chinese Academy of Sciences, Shanghai 201800, China*

[2]*Center of Materials Science and Optoelectronics Engineering，University of Chinese Academy of Sciences, Beijing 100049, China*

[3]*University of Chinese Academy of Sciences, Beijing 100049, China*

## Abstract

The generation of the below threshold harmonics (BTHs) under different driving laser intensities is investigated. The linearly shifting of photon energy and resonantly enhancement of photon yield of the harmonics from 23[rd] (H23) to 27[th] (H27) are found by changing the laser intensity around 18.6 TW/cm$^2$. It is identified that this driving laser intensity dependence is due to the transient ac Stark-shifted resonance between the first excited state and the ground state. With this transient ac Stark, the linearly shifting of the photon energy can be interpreted very well by considering the sub-cycle electron dynamics for BTH generation, which shows that the generation of the BTHs is surprisingly similar to the plateau harmonics.

High-order harmonics generated by the interaction of extremely intense laser field with noble and simple polyatomic gases have been extensively studied and utilized to produce an intense coherent XUV or X-ray light source [1-3] to synthesize isolated attosecond pulses (IAPs) or attosecond pulse trains (APTs) by synchronizing harmonics near the cutoff region [4] to probe ultrafast dynamics of tunneling ionization (TI) and rescattering of electron wave packet (EWP) of molecules and atoms with attosecond precision [5-8]. The process, which is frequently referred to as high-order harmonic generation (HHG), has been intuitively clarified by Corkum's three-step model [9]. To fully understand this process, a quantum mechanical theory by solving the time dependent Schrödinger equation (TDSE) with the strong field approximation (SFA) is also developed to precisely describe HHG [10]. In this SFA model, only the ground state is considered which means the effect of the energy structure of the atom is neglected.

So far, the effect of the energy structure is mainly investigated by numerically solving the time-dependent Schrödinger equation (TDSE). In the previous work, the resonance enhanced HHG [11-20] and the effect of the prepared excited state or the mixing state [21-24] are mainly investigated. In these works, the second order ac Stark shift is considered. With the second order ac Stark shift, the energy level will change with the laser ponderomotive energy which is proportional to the laser intensity [25, 26]. With this physical picture, the resonance enhanced HHG should satisfy the condition [12, 20]

$$\left| E_{np} - E_0 \right| + U_P = q\hbar\omega \tag{1}$$

Where $U_p$ is the pondermotive energy of the laser field, $E_{np}$ is the energy level of the $np$ state, $E_0$ is the fundamental state of the atom, $q\hbar\omega$ is the photon energy of the $q$ order harmonic. With this formula, the energy level shift is decided by the laser intensity which has no sub-cycle dynamics.

Recently, the behavior of the high order harmonics around the ionization threshold are investigated [27-32]. Although BTH generation is largely incompatible with the three-step model of HHG, the surprising result is that these harmonics are still arised from a non-perturbative process [30, 31, 33]. The XFROG characterization shows that the BTH has a clear non-perturbative negative GDD and the mechanism for BTH generation is similar to the semi-classical re-scattering process responsible for plateau harmonics. It is also found that the effects induced by the Coulomb potential also have a critical impact on these harmonics [34].

In this work, we investigate the BTH generation with the mid-infrared (MIR) laser by numerically solving the time-dependent Schrödinger equation (TDSE) in the single-electron approximation. We find that the yield and the photon energy of the BTH are greatly dependent on the laser intensity. It is identified that this dependence is due to the ac Stark-shifted induced transient energy shift between the first excited state and the ground state. With this transient ac Stark shift, we show that the behavior of the BTHs are surprisingly similar to the plateau harmonics, which also needs the three-step model [9] to shed light on the sub-cycle electron dynamics.

The one-dimensional (1D) TDSE is numerically solved for calculating the dipole oscillations generated by the MIR laser pulse [35, 36]. In this work, the soft-core potential is used to model the helium atom, $V(x) = -1/(x^2+b)^{1/2}$, where the soft-core parameter is set as $b = 0.4371$, which give the ground state energy -0.91 a.u. (atomic unit), close to that of the helium atom and $x$ is the position of the electron. The energy difference between the ground state and the first excited state is 16.16 eV. The laser field used in this work is assumed to have a constant envelop of 14 cycles with one cycle for turn-on and one cycle for turn off, $E(t) = f(t)E_0\cos(\omega t)$, where $f(t)$ is the envelope of the laser pulse, $E_0$ is the electric field of the laser pulse, $\omega = 2\pi c/\lambda$ is the angular frequency, and $\lambda$ is the central wavelength of the laser pulse. In this work, the laser wavelength is $\lambda$= 2000 nm.

In this work, we mainly focus on the BTH generation around the energy difference between the ground state and the first excited state, 16.16 eV, as shown in Fig. 1(a). In Fig. 1(a), we show the BTHs with the photon energy from 12 eV to 19 eV generated by the laser intensity from 0.2 TW/cm$^2$ to 120 TW/cm$^2$, in which the white dashed lines indicate the photon energy position of harmonics from H23 to H29.

From Fig. 1(a) we can see, until the laser intensity is larger than 90 TW/cm$^2$, the generated harmonic peaks become very clear. But there are several harmonics can be clearly seen around the photon energy of 16 eV for the laser intensity from 0.2 TW/cm$^2$ to 40 TW/cm$^2$. These harmonics appear at very low laser intensity, and their photon energy positions decrease almost linearly as the increasing of the laser intensity. To compare with the normal harmonics generated by the laser intensity higher than 90 TW/cm$^2$, we show the harmonic spectra under the laser intensity of 18.6 TW/cm$^2$ and 98 TW/cm$^2$ in Fig. 1(b). The solid-blue and solid-red lines represent the harmonics produced by the laser intensity of 18.6 TW/cm$^2$ and 98 TW/cm$^2$ respectively. For the laser intensity of 98 TW/cm$^2$, the generated harmonics are particular clear, while for the laser intensity of 18.6 TW/cm$^2$, only the harmonics located around 23$^{rd}$, 25$^{th}$ and 27$^{th}$ are well visible. Besides, for both laser intensities, the harmonics around 25$^{th}$ (H25, 15.5eV) have almost the same yield, and they are stronger than their nearby harmonics. Another feature should be noticed is that the photon energy of the harmonic around H25 from 18.6 TW/cm$^2$ is a little larger than that from 98 TW/cm$^2$. From Fig. 1(a) we can see, the yield of this harmonic first increases gradually, reaches its maximum value, and then decreases to vanishing as the laser intensity increasing. The emission is strongest when the laser intensity is about 18.6 TW/cm$^2$. When the laser intensity is approaching zero, the photon energy of this harmonic will reach 16.16 eV, which is equal to the energy difference between the ground state and the first excited state. With the increase of the laser intensity, the photon energy of this harmonic decrease almost linearly. Around the laser intensity of 18.6 TW/cm², some emissions can also be seen near H27 and H23, separated by two photon energies respectively. We believe that these harmonics generation around 18.6 TW/cm² in Fig. 1(a) is enhanced by the resonance between the ground state and the first excited state. The photon energy difference of H25 between the laser intensity of 18.6 TW/cm² and 98 TW/cm² is from the transient ac Stark shift.

In the previous work [12, 20, 37], the ac Stark shift is considered, but only the cycle-averaged shift. The multiphoton resonance with MIR laser is also investigated [12, 18, 38, 39], only the cycle-averaged ponderomotive energy proportional to the laser intensity is considered. In this work, we have to consider the transient Stark shift [37, 40]. In perturbation theory, as mentioned in [20], the second order Stark-shift can be written as below. For a certain atomic state |a> can be expressed as,

$$\Delta \varepsilon_a = \sum_{k \neq a} \frac{e^2 d_{ka}^2}{\varepsilon_k - \varepsilon_a} E^2(t) \qquad (2)$$

Where $e$ is the electron charge, E($t$) is the electric field, and $d_{ka}$ is the dipole transition element coupling the state |k> and state |a>. $\varepsilon_k$ and $\varepsilon_a$ are the energy level of quantum states |k> and |a>, respectively. Then the energy shift between the ground state and the first excited state induced by the Stark shift can be written as

$$\Delta \varepsilon(t) = \Delta \varepsilon_1 - \Delta \varepsilon_0 = \sum_{k \neq 1} \frac{e^2 d_{k1}^2}{\varepsilon_k - \varepsilon_1} E^2(t) - \sum_{k \neq 0} \frac{e^2 d_{k0}^2}{\varepsilon_k - \varepsilon_0} E^2(t)$$
$$= \left( \sum_{k \neq 1} \frac{e^2 d_{k1}^2}{\varepsilon_k - \varepsilon_1} - \sum_{k \neq 0} \frac{e^2 d_{k0}^2}{\varepsilon_k - \varepsilon_0} \right) E^2(t) = \alpha E^2(t) \qquad (3)$$

Where "0" and "1" for the ground state and 1 for the first excited state respectively. Obviously, the energy shift between two states induced by the Stark shift changes linearly with the laser electric field squared. The maximum (minimum) of the laser electric field lead to the maximum (minimum) of the Stark shift. In this work, α equals to -17.8 a.u., which is about -0.0138 eV/(TW/cm$^2$).

For the three-step model of HHG [9], the electron is mainly ionized around the crest of the electric field and recombine to the ground state after a little time, while for the generation of the below-threshold harmonics, this ionization step should be replaced by the excitation of the electron from the ground state to its excited state. So, if the electron is resonantly excited to the first excited state around the crest of the oscillating electric field where the Stark shift is largest and recombine back to the ground state after a little time, the emitted resonance harmonic (RH) photon energy divided by the laser photon energy may not be round number. It will be a little larger than that of the normal harmonics because the Stark shift will be smaller at the combination time. With this physical picture, we can do following analysis based on the transient Stark shift. We assume that the electron excitation efficiency reaches its maximum when the laser field turns to its peak, like the tunneling ionization. Therefore, the time at where the laser electric field reaches its peaks can be treated as the beginning time (excitation time) $T_b$ to generate the harmonics. In addition, at time $T_b$, if the Stark-shifted energy difference between the ground state and the first excited state is equal to the odd times of the driving laser photon energy, the excitation efficiency will be greatly enhanced by the multiphoton resonance. In Fig. 2, the Stark-shifted energy difference $\Delta\varepsilon(t)$ is plotted as solid-red curves for laser intensities of 15.5 TW/cm$^2$, 18.6 TW/cm$^2$ and 21.9 TW/cm$^2$, respectively. In the meantime, the energy position of the harmonics H23 to H27 is shown as the solid-carmine lines, and the laser field squared is plotted by the solid-blue curve. The excitation time $T_b$ is indicated by the black arrow. Only one optical cycle of the laser pulse, from -0.5 O.C. (O.C. for optical cycle) to 0.5 O.C. is shown. The excited electron will recombine to the ground state around the time $T_r$.

In Fig. 2, the Stark-shift energy difference between the ground state and the first excited state is $\Delta E = \varepsilon_1 - \varepsilon_0 + \Delta\varepsilon(t) = 16.16$ eV $+\Delta\varepsilon(t)$. Here, $\varepsilon_{0,1}$ is the free field atomic energy level, respectively. $\Delta\varepsilon(t)$ is the Stark shift calculated by equation (3). At the time t = 0.25 O.C., the laser intensities (solid red curves) from top to bottom are 15.5 TW/cm$^2$, 18.6 TW/cm$^2$ and 21.9 TW/cm$^2$, respectively. At the time $T_b$, if the laser intensity is small, e.g. 15.5 TW/cm$^2$, the Stark-shifted energy difference $\Delta E$ is far from multiphoton resonance. The excitation rate will be small. As the increasing of the laser intensity, the Stark-shifted energy difference is getting closer to the resonance. When the laser intensity is around 18.6 TW/cm$^2$, the Stark-shifted energy difference is perfectly resonant with 25 driving laser photons, resulting the highest yield of the harmonic. This agrees with the results shown in Fig. 1, where the strongest emission of the 25th harmonic around 16 eV happens at the driving laser intensity of 18.6 TW/cm$^2$. As the driving laser intensity further increases, the Stark-shifted energy difference becomes away from the resonance gradually, resulting the decrease of the 25th harmonic yield.

To quantitatively determine the photon energy shift of the released 25th harmonic with the three-step model and the time-dependent Stark shift, the recombination time $T_r$ in Fig. 2 is needed, which is a little difficult to identify. In previous work for above threshold harmonics, this recombination time Tr can be obtained from the time-frequency analysis of the time-dependent dipole oscillation [41]. But this wavelet transformation has limited energy resolution or temporal resolution, therefore the resonant effect on below threshold harmonics is difficult to be identified. In this work, we use the short time Fourier transform (STFT) [42, 43] to determine the recombination time of the harmonics generated at laser intensity of 15.5 TW/cm$^2$, as shown in Fig. 3. This STFT method has much higher temporal resolution and has successfully revealed the quantum dynamics of atomic hydrogen in intense laser field. In Fig. 3, only the results around the RHs are shown. The energy position of the RH is indicated by the dashed-white line and the Stark-shifted energy difference ΔE is shown by the solid-white curve. From the STFT result, the RH emitting all the time with a little different energy in keeping with the Stark shift. But only the photon energy marked by the dashed-white line has the constructively interference to produce an emission peak. Here we simply use the intersection point of the dashed-white line and the solid-white curve to determine the recombination time $T_r$, as marked by the dashed-black line in the figure.

So, the excitation time $T_b$ and the recombination time $T_r$ of the RHs can all be obtained now. As shown in Fig. 2, the excitation time $T_b$ is shown by the black arrow, and the recombination time $T_r$ is indicated by the dashed-black line. Around the excitation time $T_b$, the electron will be excited to the Stark shifted first excited state. With the increase of the laser intensity, the energy difference between the ground state and the first excited state will become smaller. After a while, at the recombination time $T_r$, the electron recombines back to the ground state. Because the electric field squared at this time becomes smaller than that of excitation time $T_b$, the energy difference between the ground state and the first excited state becomes larger, which will emit a photon with higher photon energy than that needed for the excitation. For example, around 18.6 TW/cm$^2$, the electron absorbs 25 driving laser photons and is multiphoton resonantly excited to the first excited state. When the electron recombines back to the ground state and emit a harmonic photon, the photon energy of the harmonic will be a little larger than the sum of 25 driving laser photons. It is obvious that the stronger the laser intensity, the lower the Stark-shifted energy, leading to smaller emitted photon energy of the RHs.

Hence, when the laser intensity is closing to zero, the emitted photon energy of the RHs will be equal to the non-Stark-shifted energy level. If we take this non-Stark-shifted energy difference as a reference, the energy shift for laser intensities from 1.7 TW/cm$^2$ to 33.7 TW/cm$^2$ in Fig. 1(a) extracted as a function of the laser intensity is plotted in Fig. 4(a), indicated by light-green-rectangle line. The Stark shift induced energy difference is calculated by the equation (3), considering the excitation time $T_b$ and the recombination time $T_r$, indicated by the black-cycle line in Fig. 4(a). The well agreement between the two curves means the photon energy shift of the harmonics (H23 to H27) can be interpreted by the transient Stark effect and the recombination time retrieved from the STFT analysis is correct. In addition, when these two curves are plotted into the Fig. 1(a), as shown in Fig. 4(b), the agreement between the transient Stark effect and the TDSE result can be clearly seen.

In conclusion, the resonantly enhanced BTHs generated by MIR laser pulse is numerically investigated. From the simulation, we found that the BTHs (from H23 to H27) yield is greatly enhanced around driving laser intensity of 18.6 TW/cm$^2$ (from 0.2 TW/cm$^2$ to 40 TW/cm$^2$) and their photon energies are strangely dependent on the laser intensity, decreasing with the laser intensity linearly. These BTHs may have been observed in previous works [44, 45], but no clear explanation is shown. With the transient ac Stark shift effect and the three-step model, we show that photon energy shifting of the BTHs can be explained quantitively very well by considering the excitation time $T_b$ and the recombination time $T_r$ retrieved from the STFT time-frequency analysis, which surprisingly indicate that the behavior of the BTHs is similar to the plateau harmonics.

We thank Professor Feng He of Jiaotong University for detail discussion for Stark shift. This work was supported by National Natural Science Foundation of China (61690223, 61521093, 11127901, 11574332, 11774363); Strategic Priority Research Program of the Chinese Academy of Sciences (Grant No. XDB16).    *: Li Wang and Jinxing Xue contributed equally to this work.         #email address: zhinan_zeng@mail.siom.ac.cn,    †email address: ruxinli@mail.shcnc.ac.cn.

**References**

1. T. Ditmire, E. T. Gumbrell, R. A. Smith, J. W. G. Tisch, D. D. Meyerhofer, and M. H. R. Hutchinson, "Spatial Coherence Measurement of Soft X-Ray Radiation Produced by High Order Harmonic Generation," Physical Review Letters 77, 4756-4759 (1996).

2. M. Bellini, C. Lyngå, A. Tozzi, M. B. Gaarde, T. W. Hänsch, A. L'Huillier, and C. G. Wahlström, "Temporal Coherence of Ultrashort High-Order Harmonic Pulses," Physical Review Letters 81, 297-300 (1998).

3. P. Salières, A. L'Huillier, and M. Lewenstein, "Coherence Control of High-Order Harmonics," Physical Review Letters 74, 3776-3779 (1995).

4. Y. Mairesse, A. de Bohan, L. J. Frasinski, H. Merdji, L. C. Dinu, P. Monchicourt, P. Breger, M. Kovačev, R. Taïeb, B. Carré, H. G. Muller, P. Agostini, and P. Salières, "Attosecond Synchronization of High-Harmonic Soft X-rays," Science 302, 1540-1543 (2003).

5. H. Niikura, H. J. Wörner, D. M. Villeneuve, and P. B. Corkum, "Probing the Spatial Structure of a Molecular Attosecond Electron Wave Packet Using Shaped Recollision Trajectories," Physical Review Letters 107, 093004 (2011).

6. M. Kitzler and M. Lezius, "Spatial Control of Recollision Wave Packets with Attosecond Precision," Physical Review Letters 95, 253001 (2005).

7. A. Baltuška, T. Udem, M. Uiberacker, M. Hentschel, E. Goulielmakis, C. Gohle, R. Holzwarth, V. S. Yakovlev, A. Scrinzi, T. W. Hänsch, and F. Krausz, "Attosecond control of electronic processes by intense light fields," Nature 421, 611 (2003).

8. D. Shafir, Y. Mairesse, D. M. Villeneuve, P. B. Corkum, and N. Dudovich, "Atomic wavefunctions probed through strong-field light–matter interaction," Nature Physics 5, 412 (2009).

9. P. B. Corkum, "Plasma perspective on strong field multiphoton ionization," Physical Review Letters 71, 1994-1997 (1993).

10. M. Lewenstein, P. Balcou, M. Y. Ivanov, A. L'Huillier, and P. B. Corkum, "Theory of high-harmonic generation by low-frequency laser fields," Physical Review A 49, 2117-2132 (1994).

11. Z. Zhinan, L. Ruxin, C. Ya, Y. Wei, and X. Zhizhan, "Resonance-Enhanced High-Order Harmonic Generation and Frequency Mixing in Two-Color Laser Field," Physica Scripta 66, 321 (2002).

12. C. F. de Morisson Faria, R. Kopold, W. Becker, and J. M. Rost, "Resonant enhancements of high-order harmonic generation," Physical Review A 65, 023404 (2002).

13. R. A. Ganeev, L. B. E. Bom, J. C. Kieffer, and T. Ozaki, "Systematic investigation of resonance-induced single-harmonic enhancement in the extreme-ultraviolet range," Physical Review A 75, 063806 (2007).

14. K. Ishikawa, "Photoemission and Ionization of He+ under Simultaneous Irradiation of Fundamental Laser and High-Order Harmonic Pulses," Physical Review Letters 91, 043002 (2003).

15. P. Ackermann, H. Münch, and T. Halfmann, "Resonantly-enhanced harmonic generation in Argon," Opt. Express 20, 13824-13832 (2012).

16. P. V. Redkin, M. B. Danailov, and R. A. Ganeev, "Endohedral fullerenes: A way to control resonant high-order harmonic generation," Physical Review A 84, 013407 (2011).

17. S. Fang, T. Tanigawa, K. L. Ishikawa, N. Karasawa, and M. Yamashita, "Isolated attosecond pulse generation by monocycle pumping: the use of a harmonic region with minimum dispersion," J. Opt. Soc. Am. B 28, 1-9 (2011).

18. M. B. Gaarde and K. J. Schafer, "Enhancement of many high-order harmonics via a single multiphoton resonance," Physical Review A 64, 013820 (2001).

19. A. Ferré, A. E. Boguslavskiy, M. Dagan, V. Blanchet, B. D. Bruner, F. Burgy, A. Camper, D. Descamps, B. Fabre, N. Fedorov, J. Gaudin, G. Geoffroy, J. Mikosch, S. Patchkovskii, S. Petit, T. Ruchon, H. Soifer, D. Staedter, I. Wilkinson, A.

Stolow, N. Dudovich, and Y. Mairesse, "Multi-channel electronic and vibrational dynamics in polyatomic resonant high-order harmonic generation," Nature Communications 6, 5952 (2015).

20. S. Camp, K. J. Schafer, and M. B. Gaarde, "Interplay between resonant enhancement and quantum path dynamics in harmonic generation in helium," Physical Review A 92, 013404 (2015).

21. P. M. Paul, T. O. Clatterbuck, C. Lyngå, P. Colosimo, L. F. DiMauro, P. Agostini, and K. C. Kulander, "Enhanced High Harmonic Generation from an Optically Prepared Excited Medium," Physical Review Letters 94, 113906 (2005).

22. X. Yuan, P. Wei, C. Liu, Z. Zeng, Y. Zheng, J. Jiang, X. Ge, and R. Li, "Enhanced high-order harmonic generation from excited argon," Applied Physics Letters 107, 041110 (2015).

23. J. B. Watson, A. Sanpera, X. Chen, and K. Burnett, "Harmonic generation from a coherent superposition of states," Physical Review A 53, R1962-R1965 (1996).

24. J. Chen, R. Wang, Z. Zhai, J. Chen, P. Fu, B. Wang, and W.-M. Liu, "Frequency-selected enhancement of high-order-harmonic generation by interference of degenerate Rydberg states in a few-cycle laser pulse," Physical Review A 86, 033417 (2012).

25. R. R. Freeman, P. H. Bucksbaum, H. Milchberg, S. Darack, D. Schumacher, and M. E. Geusic, "Above-threshold ionization with subpicosecond laser pulses," Physical Review Letters 59, 1092-1095 (1987).

26. P. Agostini, P. Breger, A. L'Huillier, H. G. Muller, G. Petite, A. Antonetti, and A. Migus, "Giant Stark shifts in multiphoton ionization," Physical Review Letters 63, 2208-2211 (1989).

27. W.-H. Xiong, J.-W. Geng, J.-Y. Tang, L.-Y. Peng, and Q. Gong, "Mechanisms of Below-Threshold Harmonic Generation in Atoms," Physical Review Letters 112, 233001 (2014).

28. V. Strelkov, "Role of Autoionizing State in Resonant High-Order Harmonic Generation and Attosecond Pulse Production," Physical Review Letters 104, 123901 (2010).

29. W.-H. Xiong, L.-Y. Peng, and Q. Gong, "Recent progress of below-threshold harmonic generation," Journal of Physics B: Atomic, Molecular and Optical Physics 50, 032001 (2017).

30. M. Chini, X. Wang, Y. Cheng, H. Wang, Y. Wu, E. Cunningham, P.-C. Li, J. Heslar, D. A. Telnov, S.-I. Chu, and Z. Chang, "Coherent phase-matched VUV generation by field-controlled bound states," Nature Photonics 8, 437 (2014).

31. D. C. Yost, T. R. Schibli, J. Ye, J. L. Tate, J. Hostetter, M. B. Gaarde, and K. J. Schafer, "Vacuum-ultraviolet frequency combs from below-threshold harmonics," Nature Physics 5, 815 (2009).

32. S. Kim, J. Jin, Y.-J. Kim, I.-Y. Park, Y. Kim, and S.-W. Kim, "High-harmonic generation by resonant plasmon field enhancement," Nature 453, 757 (2008).

33. E. P. Power, A. M. March, F. Catoire, E. Sistrunk, K. Krushelnick, P. Agostini, and L. F. DiMauro, "XFROG phase measurement of threshold harmonics in a Keldysh-scaled system," Nature Photonics 4, 352 (2010).

34. Wei-Hao Xiong, Ji-Wei Geng, Jing-Yi Tang, Liang-You Peng, and Qihuang Gong, "Mechanisms of Below-Threshold Harmonic Generation in Atoms," Physical Review Letters 112, 233001 (2014).

35. C. Liu, Z. Zeng, P. Wei, P. Liu, R. Li, and Z. Xu, "Driving-laser wavelength dependence of high-order harmonic generation in H2+ molecules," Physical Review A 81, 033426 (2010).

36. Y. Zheng, Z. Zeng, R. Li, and Z. Xu, "Isolated-attosecond-pulse generation due to the nuclear dynamics of H2+ in a multicycle midinfrared laser field," Physical Review A 85, 023410 (2012).

37. M. Chini, B. Zhao, H. Wang, Y. Cheng, S. X. Hu, and Z. Chang, "Subcycle ac Stark Shift of Helium Excited States Probed with Isolated Attosecond Pulses," Physical Review Letters 109, 073601 (2012).

38. R. A. Ganeev, Z. Wang, P. Lan, P. Lu, M. Suzuki, and H. Kuroda, "Indium plasma in single- and two-color mid-infrared fields: Enhancement of tunable harmonics," Physical Review A 93, 043848 (2016).

39. R. A. Ganeev, "Resonance processes during harmonic generation in plasmas using mid-infrared radiation," Optics and Spectroscopy 123, 117-138 (2017).

40. F. He, C. Ruiz, A. Becker, and U. Thumm, "Attosecond probing of instantaneous ac Stark shifts in helium atoms," Journal of Physics B: Atomic, Molecular and Optical Physics 44, 211001 (2011).

41. X.-M. Tong and S.-I. Chu, "Probing the spectral and temporal structures of high-order harmonic generation in intense laser pulses," Physical Review A 61, 021802 (2000).

42. Y.-l. Sheu, L.-Y. Hsu, H.-t. Wu, P.-C. Li, and S.-I. Chu, "A new time-frequency method to reveal quantum dynamics of atomic hydrogen in intense laser pulses: Synchrosqueezing transform," AIP Advances 4, 117138 (2014).

43. P.-C. Li, Y.-L. Sheu, H. Z. Jooya, X.-X. Zhou, and S.-I. Chu, "Exploration of laser-driven electron-multirescattering dynamics in high-order harmonic generation," Scientific Reports 6, 32763 (2016).

44. M. Ferray, A. L'Huillier, X. F. Li, L. A. Lomprk, G. Mainfray, and C. Manus, " Multiple-harmonic conversion of 1064 nm radiation in rare gases ", J. Phys. B 21, L31 (1988).

45. Wei-Hao Xiong, Jian-Zhao Jin, Liang-You Peng, and Qihuang Gong, " Numerical observation of two sets of low-order harmonics near the ionization threshold ", PHYSICAL REVIEW A 96, 023418 (2017).

**Figure Captions:**

FIG. 1. (a) the BTHs generated by different laser intensities. The white dashed lines represent the photon energy position of high harmonics from H23 to H29. (b) the generated high order harmonics under two laser intensities 18.6 TW/cm$^2$ (blue-solid line) and 98 TW/cm$^2$ (red-solid line).

FIG. 2. The solid red curves show the Stark-shifted energy difference for different laser intensities, and the solid carmine lines show the energy position of harmonic orders H23, H25 and H27. The solid blue curve is the laser electric field squared as a function of time.

FIG. 3. The STFT analysis of the time-dependent dipole. The dashed white line is the energy position, 15.9 eV, of the RH. The laser wavelength and the laser intensity used for the calculation of this RH are $I_0$ = 15.5 TW/cm$^2$ and $\lambda$ = 2000 nm respectively. The solid-white curve is the Stark-shifted energy difference $\Delta$E.

FIG. 4. (a) the Stark shift $\Delta\varepsilon(t)$ and the energy shift of the harmonics retrieved from Fig. 1. The black-cycle line shows the Stark shift as a function of the driving laser intensity, and the energy shift is shown by the light-green-rectangle line. (b) the Stark-shifted energy difference and the energy shift of the RHs are plotted together with the harmonics in Fig. 1(a).

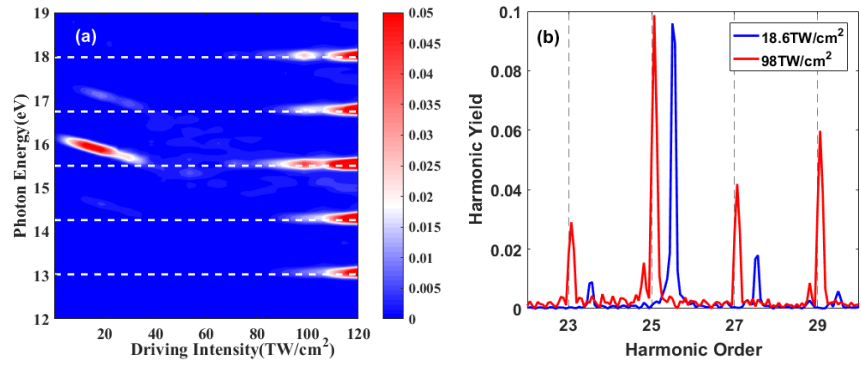

FIG. 1. (a) the BTHs generated by different laser intensities. The white dashed lines represent the photon energy position of high harmonics from H23 to H29. (b) the generated high order harmonics under two laser intensities 18.6 TW/cm$^2$ (blue-solid line) and 98 TW/cm$^2$ (red-solid line).

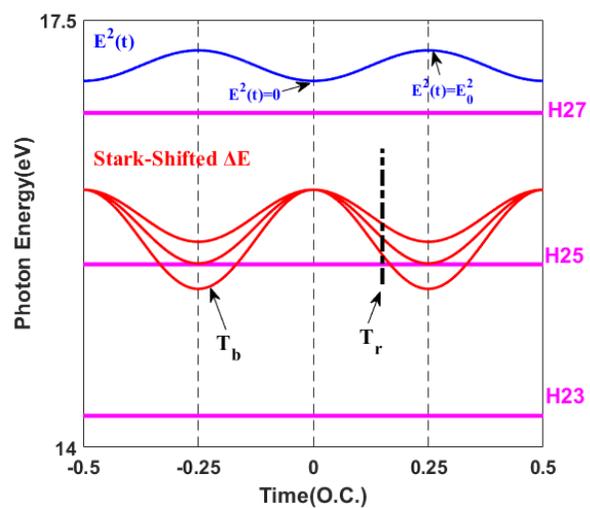

FIG. 2. The solid red curves show the Stark-shifted energy difference for different laser intensities, and the solid carmine lines show the energy position of harmonic orders H23, H25 and H27. The solid blue curve is the laser electric field squared as a function of time.

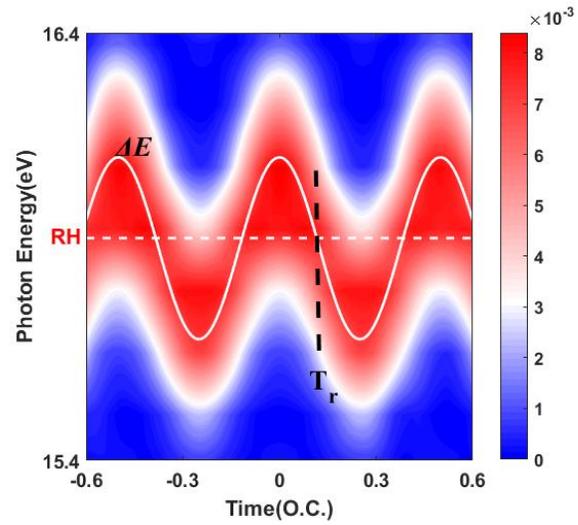

FIG. 3. The STFT analysis of the time-dependent dipole. The dashed white line is the energy position, 15.9 eV, of the RH. The laser wavelength and the laser intensity used for the calculation of this RH are $I_0 = 15.5$ TW/cm$^2$ and $\lambda = 2000$ nm respectively. The solid-white curve is the Stark-shifted energy difference $\Delta E$.

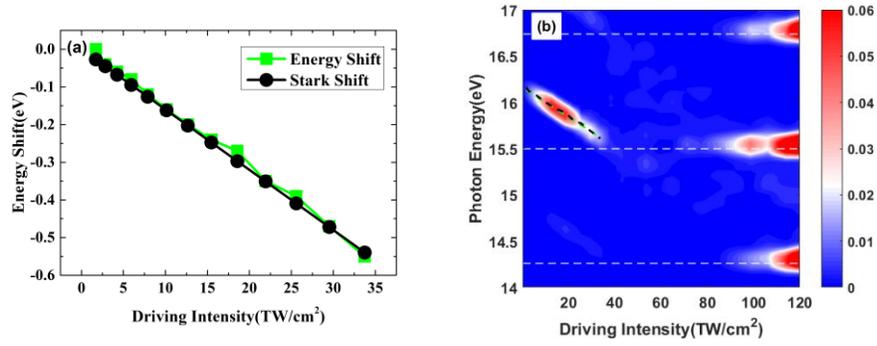

FIG. 4. (a) the Stark shift Δε(t) and the energy shift of the harmonics retrieved from Fig. 1. The black-cycle line shows the Stark shift as a function of the driving laser intensity, and the energy shift is shown by the light-green-rectangle line. (b) the Stark-shifted energy difference and the energy shift of the RHs are plotted together with the harmonics in Fig. 1(a).